\documentclass [12pt]{iopart}
\usepackage{epsf}
\eqnobysec
\begin{document}
\title{Comments on photonic shells}
\author{M \v{Z}ofka}
\address{Institute of Theoretical Physics, Faculty of Mathematics and Physics, Charles University, V Hole\v{s}ovi\v{c}k\'{a}ch 2, 180 00 Prague
8, Czech Republic} \ead{zofka@mbox.troja.mff.cuni.cz}
\begin{abstract}
We point out the similarities and differences between cylindrical
and disk photonic shells separating regions of flat spacetime.
\end{abstract}
%
%
\section*{Introduction}
We investigate in detail the special case of an infinitely thin static cylindrical shell composed of counter-rotating photons on circular geodetical paths separating two distinct parts of Minkowski spacetimes---one inside and the other outside the shell \cite{Bi_Zo}---and compare it to a static disk shell formed by null particles counter-rotating on circular geodesics within the shell located between two sections of flat spacetime \cite{Lemos}. One might ask whether the two cases are
not, in fact, merely one.
\section{The cylinder}
\begin{figure}[h]
\begin{center}
\epsfxsize=6cm 
\epsfbox{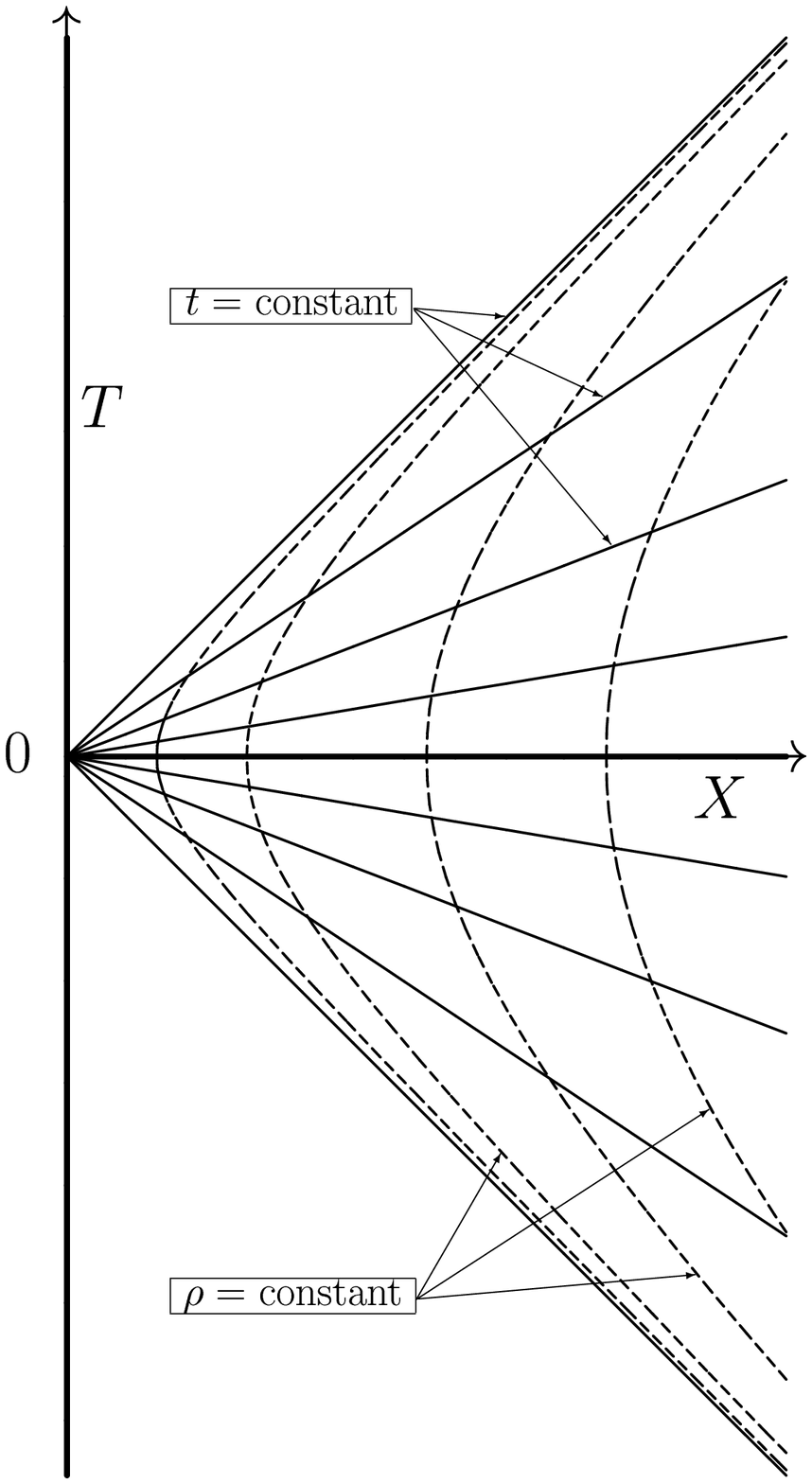}
\end{center}
\caption{\label{Figure - Rindler Coordinates} Rindler cylindrical
coordinates as functions of Minkowskian coordinates. Each point in
the diagram corresponds to a plane $Y, Z \in \!
I\hspace{-0.13cm}R$.}
\end{figure}

Inside the shell we have flat spacetime metric in cylindrical coordinates
\begin{equation}\label{Cylindrical metric}
ds^2 = -dt^2+ dz^2 + d\rho^2 + \rho^2 d\varphi^2.
\end{equation}
Outside, we again have flat spacetime but we use accelerated
Rindler coordinates:
\begin{equation} \label{Rindler metric}
ds^2 = -\rho^{2} dt^2 + dz^2 + d\rho^2 + d\varphi^2 / C^2,
\end{equation}
where $C$ determines the conicity of the metric \cite{Marder on Conicity} ($\varphi \in (0,2\pi]$). The relation of this coordinate system to the Minkowski
coordinates is as follows
\begin{equation} \label{Cylinder to Rindler}
\begin{array}{rclrcl}
T & = & \rho \sinh t, & \;\;\; Y & = & \varphi/ C,\\
X & = & \rho \cosh t, & Z & = & z.\\
\end{array}
\end{equation}
If we wish to obtain the entire covering Minkowski spacetime we
must extend the ranges of $X$ and $Y$ to the entire real axis (see
also \cite{Gautreau Hoffman}). A 3-dimensional cylinder in coordinates (\ref{Rindler metric}) with $\rho=\rho_0$ corresponds to a planar strip
$X^2-T^2=\rho_0^2$ with $Z \in \! I\hspace{-0.13cm}R, \; Y \in
[0,2\pi/C)$ and identified edges that moves at the speed $T/X$ from $X=\infty$ in
the distant past, $T \rightarrow -\infty$, comes to a stop at $X=\rho_0$ and
$T=0$ and speeds off to $X=\infty$ in the distant future, $T \rightarrow \infty$, as
follows from \Fref{Figure - Rindler Coordinates}. Both metrics (\ref{Cylindrical metric}), (\ref{Rindler metric}) are
special cases of the Levi-Civita metric \cite{Levi-Civita}.

The two separate systems are connected through a cylindrical hypersurface located at the same coordinate radius $\rho=1/C$, both inside and outside. The radial coordinate is thus continuous and the remaining coordinates are connected as follows:
\begin{equation} \label{Identification}
\fl \hspace{1cm} t_- = T = t_+ \rho = t_+ / C, \hspace{1cm} z_- = Z = z_+, \hspace{1cm} \varphi_- \rho = Y = \varphi_+ / C,
\end{equation}
where $T,Z,Y$ are coordinates on the shell itself (for clarity, coordinates inside and outside the shell are denoted by $+$ and $-$ indices, respectively). With this definition, the 3-dimensional induced metric on the cylindrical shell is identical from both sides and it is flat\footnote[1]{This ensures continuity of the first fundamental forms. To be able to identify the two surfaces globally, we require that the length of hoops $t, z, \rho=$ const be the same from both sides of the shell. With our choice of the conicity parameter, $C=1/\rho$, this condition is fulfilled.}. We find the extrinsic curvatures, $K_{AB}$, of the shell inside and outside and, using Israel formalism \cite{Israel}, we calculate the induced energy-momentum tensor $S_{AB}$ on the shell (for details, refer to a more general situation in \cite{Bi_Zo})
\begin{equation} \label{Induced Tensor}
S_{AB} = \frac{1}{8\pi\rho}
\left( \begin{array}{ccc}
1 & 0 & 0 \\
0 & 0 & 0 \\
0 & 0 & 1 \\
\end{array} \right).
\end{equation}
The resulting energy-momentum tensor is diagonal with zero trace and thus it can be interpreted as streams of counter-rotating photons. Moreover, the axial component of the energy-momentum tensor is zero and so the photons only move in the $\varphi$-direction. Since they have constant coordinate velocities, their trajectories are necessarily circular geodesics within the 3-dimensional subspace of the cylindrical shell. A similar situation with multiple photonic shells has been discussed recently in \cite{Arik Delice} (the authors use the Kasner form of the metric\footnote{The applied matching conditions are not generic here since the authors only obtain shells composed of null particles although it has been demonstrated that this is not the case in general \cite{Bi_Zo}.})---we focus on a special case of a single shell formed by non-helical photons with flat regions both inside and outside.

If we calculate the mass of the shell  per unit proper length, $M_1$, we find out $M_1 \equiv 2 \pi \rho \; S_{TT} = 1/4$. It might be surprising that this characteristic mass is not zero since there is flat spacetime both inside and outside of the shell. However, there is no single {\it global} coordinate system that would enable us to describe the spacetime using the Minkowski metric. In other words the shell {\it does} produce gravitational field since there are freely moving observers who are accelerated towards one another.

\begin{figure}[h]
\begin{center}
\epsfxsize=6cm 
\epsfbox{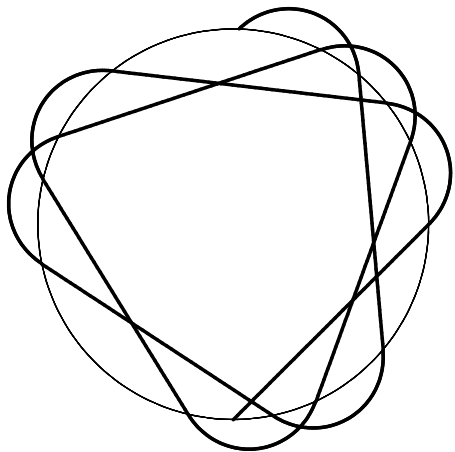}
\end{center}
\caption{\label{Trajectory-Cylindrical Coordinates} Cylindrical coordinates: A typical trajectory of a free massive particle entering and leaving the cylinder repeatedly.}
\end{figure}

\begin{figure}[h]
\begin{center}
\epsfxsize=8cm 
\epsfbox{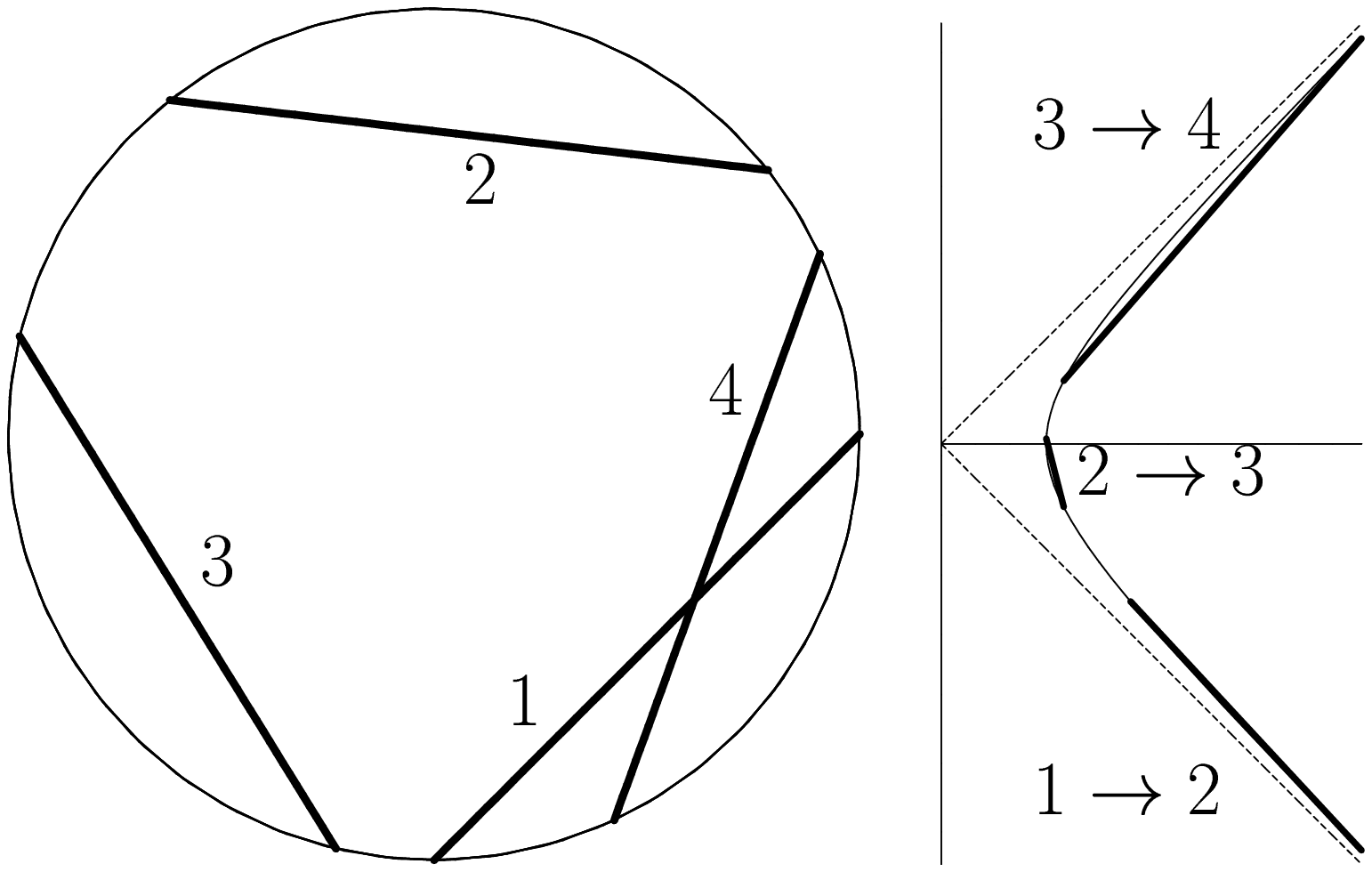}
\end{center}
\caption{\label{Trajectory-Minkowski Coordinates} The same geodesic in Minkowski coordinates: The particle begins travelling along path labelled $1$ within the cylinder, then crosses the shell and appears outside on path $1 \rightarrow 2$. After a finite time it crosses the shell again and continues along path labelled $2$ eventually hitting the shell again---and so forth.}
\end{figure}

Let us look at the geodesics of test particles passing through
the shell. Inside and outside of the cylinder they just move along straight lines
at a constant speed (if we use Minkowskian coordinates). Crossing the sheet, the projection of the
particle's 4-velocity on a chosen tangential triad and the normal
vector must be the same on both sides (we neglect here any interaction with the shell). This gives continuous
4-velocity components except for the temporal component with
\begin{equation}
U^t|_+ = U^t|_- \; / \rho.
\end{equation}
The direction of motion of the particle thus remains the same but
the magnitude of its coordinate velocity changes. Outside the
shell, we find
\begin{equation} \label{Trajectories Outside}
\begin{array}{rclrcl}
t & = & t_0 + \mbox{arctanh} \left[ \tau \frac {\sqrt {\alpha^2 + (\dot{\rho}_0)^2}} {( \rho_0 + \tau \dot{\rho}_0 )} \right], \vspace{0.2cm} \hspace{1cm} & z & = & z_0 + \dot{z}_0 \; \tau, \vspace{0.2cm} \\
\rho & = & \sqrt{ \rho_0^2 + 2 \; \tau \; \rho_0 \; \dot{\rho}_0 - \alpha^2 \tau^2 }, \vspace{0.2cm} \hspace{1cm} & \varphi & = & \varphi_0 + \dot{\varphi}_0 \; \tau,
\end{array}
\end{equation}
with $\alpha^2 = 1 + (\dot{z}_0)^2 + (\dot{\varphi}_0)^2 / C^2 \geq 1$ for massive particles and $\alpha^2 = (\dot{z}_0)^2 + (\dot{\varphi}_0)^2 / C^2 \geq 0$ for photons, where $\tau$ is the proper time and affine parameter, respectively. The symbols $t_0, \rho_0, z_0, \varphi_0,
\dot{\rho}_0, \dot{z}_0$ and $\dot{\varphi}_0$ stand for the values of the
coordinate time, distance from the axis, axial and angular
coordinates of the particle, and its radial, axial and angular
velocities at $\tau=0$\footnote{This is of course a straight-line
geodesic in the covering Minkowski spacetime.}. Free massive particles and photons
leaving the cylinder fall back after a finite proper time/affine parameter interval under
the same angle they left the shell before. After crossing the shell, they thus circumscribe always the same trajectory outside. An example of a typical trajectory in cylindrical coordinates is shown in \Fref{Trajectory-Cylindrical Coordinates}. Only radially moving photons escape to infinity. This can be easily seen in \Fref{Trajectory-Minkowski Coordinates} that uses Minkowskian coordinates both inside and outside of the shell---all geodesics (straight lines here) that leave the shell at some point and lie within the light cone intersect the
cylindrical hypersurface again. Any observer outside perceives the
cylinder as an infinite planar wall falling upon him
at the speed $T/X$. No matter how hard he tries, the wall
always hits him. After
he passes through the shell, he emerges in another section of
Minkowski spacetime. He realizes he is surrounded by a cylinder.
Using a rocket, he is able to stay in this part of the spacetime
indefinitely. However, he can always penetrate the wall again and
reemerge in his original universe.
\section{The disk}
The metric within the flat spacetime regions above and below the
disk surface reads
\begin{equation}\label{Disk coordinates}
ds^2 = - \zeta^2 d (\tau / 2)^2 + d\zeta^2 + d\eta^2 + \eta^2
d\phi^2.
\end{equation}
Using the following transformation
\begin{equation}
\begin{array}{rclrcl}
T & = & \zeta \sinh (\tau /2), \hspace{1cm} & Y & = & \eta \cos \phi, \\
X & = & \zeta \cosh (\tau /2), \hspace{1cm} & Z & = & \eta \sin \phi,
\end{array}
\end{equation}
we find the usual Minkowski metric. An observer located
above and below the disk sees it as a hypersurface defined by $X =
\pm \sqrt { T^2 + Z^2 + Y^2 } $, respectively. Identifying these,
we find out that the disk is composed of counter rotating null
particles moving on circular geodesics around the disk centre. The surface density is positive definite everywhere. Any massive geodetical test particle penetrating the disk spirals down towards
the centre of the disk in infinite proper time while photons are able to escape to
infinity. A very detailed discussion is to be found in
\cite{Lemos}.
\section{Comparison}
The similarity between these solutions lies in the fact that a thin shell of matter separates two regions of flat spacetime and that the shell can be interpreted as free photons moving along circular orbits within the shell. One might then ask what holds the shell in place. From the point of view of the coordinate systems (\ref{Cylindrical metric}) and (\ref{Rindler metric}), where the shell surface is static, the centrifugal force acting upon the shell particles is balanced by the gravitational force exerted upon them by the other particles on the shell. If we adopt the perspective of two Minkowskian coordinate systems (\ref{Cylindrical metric}) and (\ref{Cylinder to Rindler}) then, however, the outer shell surface is not static. It is, in fact, accelerated towards free particles at rest in such a system. This effect balances the centrifugal force acting on the shell.

The difference between the two cases consists in the definition of the hypersurfaces that are identified in the two separate Minkowski spacetimes inside and outside, and above and below, respectively. If we want to compare these hypersurfaces, we need to use a single coordinate system. We begin by describing the outer cylindrical hypersurface with a coordinate radius $\rho$ in the disk coordinates (\ref{Disk coordinates}). We first transform the coordinate system (\ref{Rindler metric}) into the Minkowskian coordinates and from here we go to (\ref{Disk coordinates}) to obtain
\begin{equation}\label{Rindler - Cylinder - Outer}
\zeta_+ = \rho.
\end{equation}
This is just a planar horizontal section of the spacetime. The inner hypersurface, as defined in coordinates (\ref{Cylindrical metric}), reads
\begin{equation}\label{Rindler - Cylinder - Inner}
\eta_- = \rho.
\end{equation}
In this case, the cylindrical surface is again a cylinder. We identify points with
\begin{equation} \label{Identification in Rindler - Cylinder}
\begin{array}{rclrcl}
\zeta_- & = & \sqrt{\eta_+^2 \; \sin^2 \phi_+ - \frac {\tau_+^2} {4} \rho^2}, \hspace{1cm} & \eta_- & = & \zeta_+ = \rho, \nonumber \\\\
\tau_- & = & 2 \; \mbox{arctanh} \frac {\tau_+ \; \rho} {2 \eta_+ \; \sin \phi_+}, \hspace{1cm} & \phi_- & = & \frac {1} {\rho} \eta_+ \cos \phi_+,
\end{array}
\end{equation}
where the minus sign refers to the inner (lower) hypersurface and plus to the outer (upper) one. The identification is not at all trivial and, moreover, it is time-dependent.

In coordinates defined by (\ref{Disk coordinates}), the disk surface is given by the cones
\begin{equation}\label{Rindler - Disk}
\zeta_\pm = \pm \eta_\pm,
\end{equation}
where we identify points with
\begin{equation} \label{Identification in Rindler - Disk}
\begin{array}{rclrcl}
\zeta_- & = & -\zeta_+, \hspace{1cm} & \eta_- & = & \eta_+,\\
\tau_- & = & \tau_+,  \hspace{1cm} & \phi_- & = & \phi_+.
\end{array}
\end{equation}
It can be seen immediately that (\ref{Rindler - Cylinder - Inner}), (\ref{Rindler - Cylinder - Outer}), (\ref{Identification in Rindler - Cylinder}) and (\ref{Rindler - Disk}), (\ref{Identification in Rindler - Disk}) are not identical.
\section{Conclusions}
We conclude that the two cases are different although constructed in an analogous way using two separate Minkowskian regions. One can imagine various pairs of identification surfaces, both static and time-dependent, that produce infinitely thin shells separating regions of flat spacetime and that are composed of photons (another example is a thin spherical shell separating two identical regions of Minkowski spacetime of finite volume, see \cite{Langer Zofka}). There are only two conditions---the induced metric must be the same on both hypersurfaces and the trace of the induced energy-momentum tensor (satisfying energy conditions) needs to be zero. If we use a coordinate system in which the shell is static and which covers both the hypersurface and the neighbouring region, geodesics will lead escaping particles back towards the shell. If we, on the other hand, prefer a simpler Minkowskian description, then test particles outside are not accelerated but the shell itself speeds towards them. In either case, the existence of the shells is enabled by the resulting balance.
\ack
I would like to thank prof. J.~Bi\v{c}\'ak for discussions and prof. W.B.~Bonnor for drawing my attention to this topic.
\section*{References}\label{References}
\addtocontents{toc}{\contentsline {section}{\numberline
{}\hspace{-0.75cm} References}{\pageref{References}}}

\end{document}